\documentclass[apj,numberedappendix]{emulateapj}
\usepackage{apjfonts}

\newcommand{\be}{\begin{eqnarray}}
\newcommand{\ee}{\end{eqnarray}}

\newcommand{\carbon}{$^{12}$C}

\newcommand{\tdyn}{t_{\rm dyn}}

\begin{document}

\normalsize

% -----------------------------------------------------------
% -----------------------------------------------------------

\slugcomment{Accepted for publication in The Astrophysical Journal}
\shorttitle{NEUTRONIZATION DURING SN Ia SIMMERING} 
\shortauthors{PIRO \& BILDSTEN}

\title{Neutronization During Type Ia Supernova Simmering}

% -----------------------------------------------------------
% -----------------------------------------------------------

\author{Anthony L. Piro\altaffilmark{1} and Lars Bildsten} 

\affil{Kavli Institute for Theoretical Physics, Kohn Hall, University
of California, Santa Barbara, CA 93106;\\ piro@kitp.ucsb.edu, bildsten@kitp.ucsb.edu}

\altaffiltext{1}{Current address: Astronomy Department and Theoretical Astrophysics Center,
University of California, Berkeley, CA 94720; tpiro@astro.berkeley.edu}

% -----------------------------------------------------------
% -----------------------------------------------------------

\begin{abstract}

 Prior to the  incineration of a white dwarf (WD) that makes a Type Ia supernova (SN Ia), the 
star ``simmers'' for $\sim 1000$ years in a convecting, carbon burning region.
We have found that weak interactions during this time increase the 
neutron excess by an amount that depends on the total quantity of carbon burned prior to the 
explosion. This contribution is in addition to the metallicity ($Z$) dependent neutronization
through the $^{22}$Ne abundance (as studied by Timmes, Brown, \& Truran).
The main consequence is that we expect a ``floor'' 
to the level of neutronization that dominates over the metallicity contribution when
$Z/Z_\odot\lesssim2/3$, and it can be important for even larger metallicities
if substantial energy is lost to neutrinos via the convective Urca process.
This would mask any correlations between SN Ia properties and
galactic environments at low metallicities. In addition, we show that recent observations of
the dependences of SNe Ia on galactic environments make it clear that metallicity alone cannot
provide for the full observed diversity of events.

\end{abstract}


% -----------------------------------------------------------
% -----------------------------------------------------------

\section{Introduction}
\label{sec:introduction}

The use of Type Ia supernovae (SNe Ia) as cosmological distance indicators
has intensified the need to understand white dwarf (WD) explosions. Of particular
importance is the origin of the Phillips relation \citep{phi99}, an essential luminosity calibrator.
Recent models demonstrate that it can be explained by large 
variations in the abundance of stable iron group elements \citep{woo07} with
the dominant cause for  diversity likely residing in the explosion mechanism
\citep{maz07}.

   One additional variable is the metallicity of the WD core, 
which yields excess neutrons relative to protons due to the isotope $^{22}$Ne.
This is usually expressed as
\be
	Y_e = \sum_i \frac{Z_i}{A_i}X_i,
\ee
where $A_i$ and $Z_i$ are the nucleon number and charge of species $i$
with mass fraction $X_i$.
The neutronization is critical for setting the
production of the neutron-rich isotopes \citep{thi86}.
If no weak interactions occur during the explosion,
the mass fraction of $^{56}$Ni produced is simply
$X(^{56}{\rm Ni}) = 58Y_e-28$, assuming $^{56}$Ni and $^{58}$Ni are
the only burning products \citep{tbt03}. The neutronization also affects
the explosive burning, including the laminar flame speed \citep{cha07}.
However, the metallicity range of
progenitors is not large enough to account for the full SNe Ia diversity
(see \S 4), making it critical to explore all factors that determine $Y_e$. 

A potential neutronization site is the convective carbon burning core that is
active for $\sim1000\ {\rm years}$ prior to the explosion. The hydrostatic evolution
associated with this simmering phase terminates when the core temperature is
sufficiently high that burning is dynamical \citep{nom84,ww86,woo04,ww04,kwg06},
and a flame commences \citep{tw92}. We show here that protons 
from the $^{12}$C($^{12}$C,$p$)$^{23}$Na reaction during simmering capture on
$^{12}$C, and that subsequent electron captures on $^{13}$N and $^{23}$Na
decrease $Y_e$. This process continues until either the explosion occurs or 
sufficient heavy elements have been synthesized that they capture the protons instead.

In \S \ref{sec:rates}, we present simmering WD core models and summarize
the reaction chains that alter $Y_e$. We find that one proton is converted to
a neutron for each six $^{12}$C nuclei consumed for  burning at
$\rho<1.7\times 10^9\ {\rm g \ cm^{-3}}$. At densities above this, an additional
conversion occurs from an electron capture on  $^{23}$Na. Hence, the $Y_e$
in the core depends on the amount of carbon burned during simmering
and the density at which it occurs, which we quantify in \S 3. We find that
neutronization during simmering  dominates for metallicities $Z/Z_\odot\lesssim2/3$.
%If he convective Urca process is active (Lesaffre et al. 2005, and references therein),
%the additional burning required
%to become dynamicall further decreases $Y_e$. 
 We conclude in \S \ref{sec:conclusion} by discussing the
 observations and noting where future work is needed. 

%We show that 
%nuclear reactions which take place during this time could in principal be important for determining the
%neutronization of the progenitor. 
%\citet{pod06} claimed that the neutronization
%due to $^{22}$Ne is magnified if the $^{22}$Ne is able to capture protons
%This process would provide an additional multiplicative factor when
%converting from metallicity of the progenitor to neutronization, so that
%the range of stable iron group elements produced is much wider.

%This is not that small, so the signature of
%this neutronization may appear in surveys that attempt to correlate Type Ia supernovae
%properties with the host metallicities. 

% -----------------------------------------------------------
% -----------------------------------------------------------

\section{Neutron Production During Simmering}
\label{sec:rates}
 
  Thermally unstable burning begins when the energy generation rate
from carbon fusion, $\epsilon$,  exceeds neutrino cooling 
\citep{nom84}. The thin solid lines in Figure \ref{fig:simmering} show the range of ignition
curves for $X(^{12}$C$)=X(^{16}$O$)$ \citep{yak06} with the middle
line the nominal  current best.   The carbon fuses via
$^{12}$C$(^{12}$C$, p)^{23}$Na and $^{12}$C$(^{12}$C$,\alpha)^{20}$Ne with
branching ratios of $0.44$ and $0.56$, respectively.
At ``early'' times the liberated protons capture onto $^{12}$C,
while at ``late'' times enough heavy elements ($^{23}$Na or $^{23}$Ne) have
been produced that they capture the protons instead. 

    We treat the evolution during the  simmering phase as a series of
hydrostatic models consisting of an adiabatic convective core and an
isothermal surface at $10^8\ {\rm K}$. As long as the convection zone
is well described as an adiabat this is sufficient for resolving the thermal
structure without the need to explicitly solve the energy transfer equation. These
assumptions become weaker once the central temperature is
$T_c\gtrsim7\times10^8\ {\rm K}$, so that burning occurs sufficiently
quickly that there is considerable energy generation within a convective
eddy overturn timescale \citep{gw95}. The energy generation does come into
play because it sets the heating timescale, $t_h\equiv c_pT_c/\epsilon$, where $c_p$ is the specific 
heat of the liquid ions (we use linear mixing and the Coulomb energy 
from Chabrier \& Potekhin 1998), nearly given by the classical Dulong-Petit law
$c_p\approx 3k_B/\mu_i m_p$, where  $\mu_i$ is the ion mean molecular weight.
Since we evaluate $t_h$ using the central conditions it
is a lower limit since it should include the entire heat capacity of the convective region
(Piro \& Chang 2007; see related discussion for neutron stars in Weinberg et al. 2006).
In this way, for a given thermal profile there is a well-defined heating timescale,
which connects our stationary models to the true time evolution.
The thick dashed lines in Figure \ref{fig:simmering} trace out the trajectory of
the central temperature, $T_c$, and density, $\rho_c$, for $M=1.35M_\odot$ and
$M=1.37M_\odot$ (left and right, respectively), both using compositions
of $X(^{12}$C$)=0.5$, $X(^{16}$O$)=0.48$, and $X(^{22}$Ne$)=0.02$. 
These indicate that $\rho_c$ decreases with increasing $T_c$
\citep{les06,pir07}. The thick solid lines
show thermal profiles near the end of the simmering.

   The simmering phase
ends when sub-sonic convection can no longer transport the heat outwards
because the timescale of heating is now less than the convective overturn
timescale.
Since the overturn timescale depends on the integrated energy generation rate
near the WD center (while we desire a local measure of when convection
should end for simplicity), we assume that
this occurs when $t_h\sim\tdyn\equiv(G\rho_c)^{-1/2}$, the dynamical timescale.
This gives reasonable agreement to other more careful
calculations that find the simmering phase ends when $T_c\approx8\times10^8\ {\rm K}$
\citep{woo04}.
We plot $t_h=10\tdyn$ as a dotted line in Figure \ref{fig:simmering} to indicate where
simmering ends, since 
the strong temperature sensitivity of \carbon\ fusion makes this line rather
insensitive to the choice of prefactor. If the simmering phase ends earlier, it can
be considered in the context of our models by just truncating our results at a slightly
lower $T_c$.

%During this phase, the composition is mostly $^{12}$C and $^{16}$O, with a metallicity dependent 
%abundance of $^{22}$Ne.    said below.. 

\subsection{ Main Reaction Cycle at Early Times}

 At early times, only $^{12}$C, $^{16}$O, or $^{22}$Ne are potential 
proton capture nuclei. We compared these rates using
\citet{cf88}, including strong screening \citep{svh69}. The
$^{16}$O$(p,\gamma)^{17}$F reaction is negligible, whereas resonances in 
the $^{22}$Ne$(p,\gamma)^{23}$Na reaction make its rate 
comparable to $^{12}$C$(p,\gamma)^{13}$N. However, the larger abundance of 
$^{12}$C means that it captures more protons
by a factor of $(22/12)X(^{12}$C$)/X(^{22}$Ne$)\approx40$. The fate of the synthesized
$^{13}$N  requires some discussion, as the branching amongst the three relevant
reactions: $^{13}$N$(e^-,\nu_e)^{13}$C, 
$^{13}$N$(\gamma, p)^{12}$C,  $^{13}$N$(p, \gamma)^{14}$O, 
depends on $T$, $\rho$, and proton mass fraction, $X_p$.

\begin{figure}
\epsscale{1.2}
\plotone{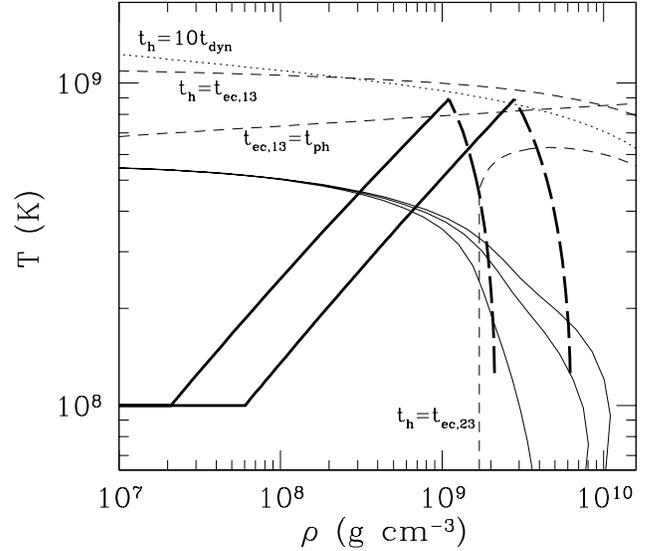}
\caption{Conditions in the simmering WD. Thin solid lines are the range of carbon ignition
curves from Yakovlev et al. (2006). The arching thick dashed lines show the central
trajectory for the simmering core of a $1.35M_\odot$ and a $1.37M_\odot$ WD
(left and right, respectively).
Thick solid lines are example thermal profiles that are composed of an adiabatic convective
core connected to an isothermal exterior. Each case is taken near the end of the simmering
phase, which occurs at the dotted line labeled $t_h=10\tdyn$.
Dashed lines are crossing points for nuclear reactions described in the text.}
\label{fig:simmering}
\epsscale{1.0}
\end{figure}

\begin{figure}
\epsscale{1.2}
\plotone{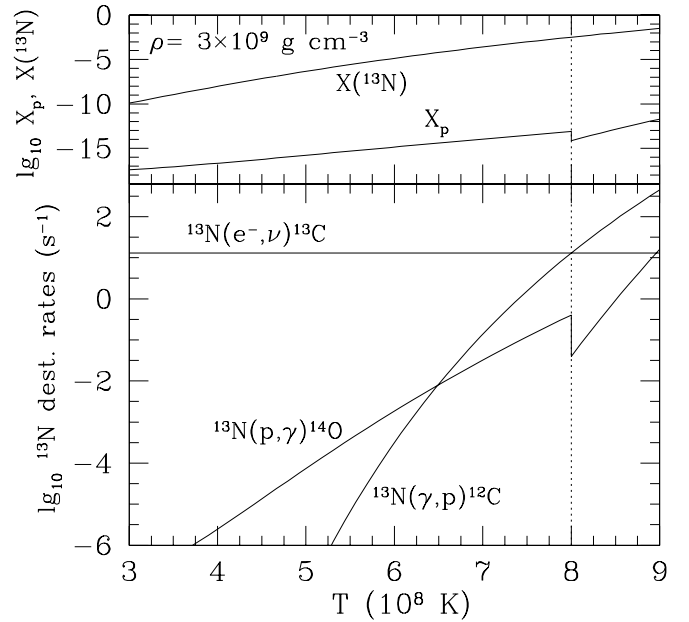}
\caption{Mass fractions $X_p$ and $X(^{13}$N) ({\it upper panel}) and
$^{13}$N destruction rates ({\it lower panel}), versus temperature.
The dotted line at $T=8\times10^8\ {\rm K}$ divides
where $^{13}$N photodisintegrations becomes important.
The density is $3\times10^9\ {\rm g\ cm^{-3}}$ and $X(^{12}{\rm C})=0.5$. }
\label{fig:abun}
\epsscale{1.0}
\end{figure}

The production of protons is always the rate limiting step, so that each
proton is almost immediately captured by $^{12}$C.
This means that we can find $X_p$ by balancing the proton production rate
from carbon fusion, $\lambda n_{12}^2 \langle \sigma v\rangle_{12+12}$ 
(where $\lambda=0.44$ is the branching ratio for the reaction $^{12}$C$(^{12}$C$,p)^{23}$Na
and $n_{12}$ is the $^{12}$C number density) with the proton capture rate, 
$n_pn_{12}\langle \sigma v\rangle_{p+12}$, where $n_p$ is the proton
number density,
%\be
%	\frac{dn_p}{dt} = \lambda n_{12}^2\langle \sigma v\rangle_{12+12}
%		-n_pn_{12}\langle \sigma v\rangle_{p+12}\approx0,
% \ee
\be
	X_p = \lambda\frac{X(^{12}{\rm C})}{12}
		\frac{\langle \sigma v\rangle_{12+12}}{\langle \sigma v\rangle_{p+12}}.
	\label{eq:xp}
\ee
This is plotted in the upper panel of Figure \ref{fig:abun}. The small value of
$X_p$ confirms our equilibrium assumption for the proton abundance, and allows us to show 
that the  $^{13}$N$(p, \gamma)^{14}$O reaction is negligible (bottom panel of Fig.
\ref{fig:abun}) in comparison to electron captures. The bottom panel also shows the rates for
the reactions $^{13}$N$(e^-,\nu_e)^{13}$C and $^{13}$N$(\gamma, p)^{12}$C,
making it clear that electron capture dominates for $T<8 \times 10^8 \ {\rm K}$.

All of the $^{13}$N comes from the protons synthesized by carbon fusion, therefore the $^{13}$N
production rate is equal to that of protons. Once again, this is the slowest step so that
we can find the $^{13}$N abundance by balancing
$n_pn_{12}\langle \sigma v\rangle_{p+12}=R_{\rm ec}(^{13}{\rm N})n_{13}$, where
$R_{\rm ec}(^{13}$N$)$ is the  $^{13}$N electron capture rate.
Captures into excited states are unlikely to be dominant, allowing us to
use the measured $ft=4.65\times10^3\ {\rm s}$ from the ground-state transitions.
(Likewise, we use $ft=1.91\times10^5\ {\rm s}$ for
$^{23}$Na electron captures in \S 2.2.)
Combining with equation (\ref{eq:xp}) gives 
\be
	X(^{13}{\rm N}) = \lambda
		\frac{13}{12^2}\left[X(^{12}{\rm C})\right]^2
		\frac{\rho N_{\rm A}\langle\sigma v \rangle_{12+12}}{R_{\rm ec}(^{13}{\rm N})},
	\label{eq:n13}
\ee
where $N_{\rm A}$ is Avagodro's number, which is shown in the top panel of
Figure \ref{fig:abun}.  The network is completed by 
$^{13}$C$(\alpha,n)^{16}$O and $^{12}$C$(n,\gamma)^{13}$C, leading to 
a composition of one each of  $^{13}$C, $^{16}$O, $^{20}$Ne, $^{23}$Na.

%This electron capture rate is
%\be
%	R_{ec}(^{13}{\rm N})=\lp\frac{\ln 2}{ft_{13}} \rp I(^{13}{\rm N}),
%	\label{eq:ec}
%\ee
%where $ft_{13}=4.65\times10^3\ {\rm s}$ and
%$I(^{13}{\rm N})$ is the dimensionless phase space integral over the electron energy $E$,
%which for an exoergic capture is given as  \citep{lan80}
%\be
%	I(^{13}{\rm N}) = \frac{1}{(m_ec^2)^5}
%		\int _{m_ec^2}^\infty
%			\frac{E(E^2-m_e^2c^4)^{1/2}(E+Q_{13})^2dE}{1+\exp[(E-E_{\rm F})/k_{\rm B}T]},
%\ee
%where $Q_{13}=2.2205\ {\rm MeV}$ is the atomic mass excess difference between $^{13}$N and
%$^{13}$C and $E_{\rm F}$ is the Fermi energy given by \citep{bc95}
%\be
%	E_{\rm F} = 4.08\ {\rm MeV}\lp \frac{\rho_9}{\mu_e/2}\rp^{1/3}
%\ee
%(for which the electron rest mass can be neglected
%for the high densities we consider).
%The resulting electron capture rate as a function of density is plotted in
%Figure \ref{fig:n13na23} ({\it solid line}) for $T=8\times10^8\ {\rm K}$. This is largely
% independent of $T$ since $k_{\rm B}T\ll E_{\rm F}$.

There are two complications. The first 
is at high $T$'s where photodisintegration of $^{13}$N happens faster than the electron 
captures (above the dashed curve labeled by $t_{\rm ec,13}=t_{\rm ph}$
in Fig. 1). Chemical balance ($p+^{12}$C$\leftrightarrow^{13}$N$+\gamma$) 
is achieved in this limit, fixing the proton to $^{13}$N ratio. The 
$^{13}$N is then slowly removed due to electron captures.
The electron captures must always balance the proton production, 
so the $^{13}$N abundance remains identical to equation (\ref{eq:n13}). Hence, 
photodisintegration adds steps to the reaction chain (and alters the proton
density; top panel of Fig. 2) but does not modify the conclusion
that all protons released in \carbon\ burning lead to $^{13}$N
electron capture.

The second complication is the reaction $^{23}$Na$(e^-,\nu_e)^{23}$Ne
at high densities ($\rho> 1.7\times10^9\ {\rm g\ cm^{-3}} $).
This occurs to the right of the  dashed line labeled as $t_h=t_{\rm ec,23}$
in Figure~\ref{fig:simmering},
%The rate for
%this has the same form as equation (\ref{eq:ec}), but with
%$ft_{23}=1.91\times10^5\ {\rm s}$ and the space integral is given by
%\be
%	I(^{23}{\rm Na}) = \frac{1}{(m_ec^2)^5}
%		\int _{Q_{23}}^\infty
%			\frac{E(E^2-m_e^2c^4)^{1/2}(E-Q_{23})^2dE}{1+\exp[(E-E_{\rm F})/k_{\rm B}T]},
%\ee
%where $Q_{23}=4.89\ {\rm MeV}$. 
illustrating that these electron captures only take place at
certain times during the simmering phase, which we account for
in \S 3. Electron captures on $^{13}$N would not have time to occur above the dashed
line labeled $t_h=t_{\rm ec,13}$ in Figure~\ref{fig:simmering}, but this is always after the explosion. 

  The main reaction cycle is summarized in Table 1.
Six \carbon\ are consumed, producing $^{13}$C, $^{16}$O, $^{20}$Ne,
and depending on $t_h$, either $^{23}$Ne or $^{23}$Na. Therefore, during
each cycle either one or two protons have been converted to neutrons.

\begin{deluxetable}{c c}
  \tablecolumns{2} \tablewidth{0pt}
  \tablecaption{Reaction Summary for Main Cycle}

  \tablehead{
    \colhead{Reaction} & \colhead{Thermal Energy Release (MeV)} }
  \startdata
  $^{12}$C$(^{12}$C$,p)^{23}$Na & 2.239 \\
  $^{12}$C$(^{12}$C$,\alpha)^{20}$Ne & 4.617 \\
  $^{12}$C$(p,\gamma)^{13}$N & 1.944 \\
  $^{13}$N$(e^-,\nu_e)^{13}$C & 0 \\
  $^{13}$C$(\alpha,n)^{16}$O & 2.214 \\
  $^{12}$C$(n,\gamma)^{13}$C & 4.947 \\
  $^{23}$Na$(e^-,\nu_e)^{23}$Ne\tablenotemark{a} & 0
   \enddata
   \tablenotetext{a}{This reaction only occurs when $t_h>t_{\rm ec,23}$
   (Fig.~\ref{fig:simmering}).}
\end{deluxetable}

\subsection{Late Time Truncation of Neutron Production}

  The carbon burning ashes eventually become abundant enough to compete with 
$^{12}$C for proton captures.  The relevant  products are $^{20}$Ne, and either
$^{23}$Na or $^{23}$Ne. The $^{20}$Ne reactions of $^{20}$Ne$(p,\gamma)^{21}$Na
and $^{20}$Ne$(p,\alpha)^{17}$F are negligible, so we focus on
$^{23}$Na and $^{23}$Ne.

\begin{figure}
\epsscale{1.2}
\plotone{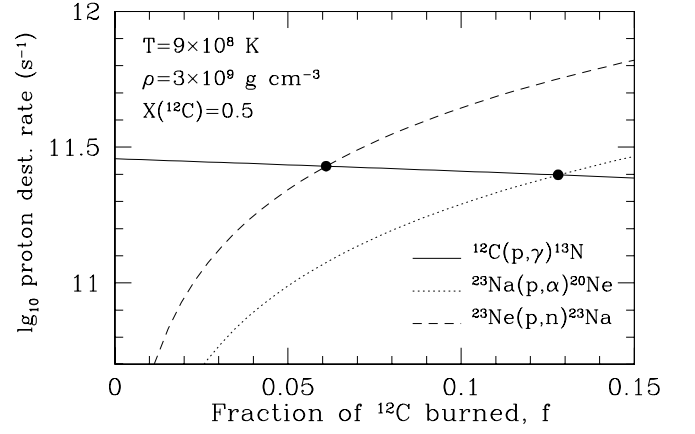}
\caption{The proton destruction rates for different processes as a function of
the fraction of $^{12}$C that has burned, $f$. The number density of $^{23}$Na
or $^{23}$Ne is taken to be $fn_{12}/3$.
Circles denote crucial places where the proton destruction mechanism switches. }
\label{fig:late}
\epsscale{1.0}
\end{figure}

We plot the proton destruction rate for the most relevant reactions in Figure~\ref{fig:late}
as a function of $f$, the fraction of $^{12}$C that has burned.
%We assume that  $T=9\times10^8\ {\rm K}$, $\rho=3\times10^9\ {\rm g\ cm^{-3}}$,
%and $X(^{12}$C$)=0.5$.
The number density of either $^{23}$Na or $^{23}$Ne is taken
to be equal to the number density of $^{12}$C burned times
$2/3\times\lambda\approx1/3$. Circles denote where the rates
cross each other, which is nearly independent of $\rho$.
For $t_h<t_{\rm ec,23}$, $f$ must exceed $\approx0.13$ 
before the $^{23}$Na$(p,\alpha)^{20}$Ne reaction becomes important, and ends 
neutronization. 

When $t_h>t_{\rm ec,23}$, $^{23}$Ne forms, which has a higher
proton capture cross section than $^{23}$Na. Burning only 
$f=0.061$ is enough that $^{23}$Ne$(p,n)^{23}$Na becomes
the primary proton sink. Since this makes $^{23}$Na and liberates a neutron,
the new $^{23}$Na may electron capture again to make $^{23}$Ne, and the reaction chain
$^{23}$Ne$(p,n)^{23}$Na$(e,\nu_e)^{23}$Ne can repeatedly occur, making many
free neutrons. Competing with this process are other
reaction chains that burn these neutrons
\citep[for example, see Table 6 of][]{at85}. In this regime, it is difficult for us to
estimate all the key nuclear reactions that will take place. Although
further neutronization is possible, we cannot follow this without
a full reaction network \citep{cha07b}. Such calculations must also be
coupled to a realistic model for the core temperature evolution (such as
what we present here).

% -----------------------------------------------------------
% -----------------------------------------------------------

\section{Maximum Neutronization Estimates}
\label{sec:models}

 We set $\eta$ as the number of protons that are converted to neutrons
for every six \carbon\ consumed, so that $\eta=2$ ($\eta=1$)
for $t_h>t_{\rm ec,23}$ ($t_h<t_{\rm ec,23}$ ), where we
approximate $\lambda\approx0.5$. The $\eta=2$ case is
an upper limit since in parts of the convection zone where $\rho<1.7\times10^9\ {\rm g\ cm^{-3}}$
the $^{23}$Na does not electron capture (and $^{23}$Ne that
is mixed to lower densities by convection may decay).
The total neutronization is measured via
\be
	Y_e= \frac{1}{2}-\frac{2X(^{56}{\rm Fe})}{56} - \frac{X(^{22}{\rm Ne})}{22}
		- f\frac{\eta X(^{12}{\rm C})}{6\times12},
\ee
which includes the initial $^{56}$Fe and $^{22}$Ne content.
Neutronization halts either when the WD explodes or when
freshly made heavy elements compete for protons (Fig.~3).

In the case of competition from fresh heavy elements,
truncation at high densities occurs when
$f=0.061$ with $\eta=2$. The maximum
change in $Y_e$ is therefore
\be
	\Delta Y_{e,\rm max} = -8.5\times10^{-4}\frac{X(^{12}{\rm C})}{0.5}.
	\label{eq:yemax}
\ee
A similar limit pertains at lower densities. One
way to exceed this limit in the high density case is if additional
reaction chains occur (see \S 2.2). We show $\Delta Y_{e,\rm max}$
as a dot-dashed line in Figure 4, in comparison to the
$\Delta Y_e$'s that result from $X(^{22}$Ne$)=0.007$ and $0.02$
({\it dotted lines}). By coincidence, the maximum effect of neutronization during
simmering is comparable to that associated with a solar metallicity.

% For simmering to contribute
%to $Y_e$ at a level comparable to that from $^{22}$Ne, the
%amount of carbon burned would be
%\be
%	f_c=0.066\ \frac{X(^{22}{\rm Ne})/0.02}{X(^{12}{\rm C})/0.5}.
%	\label{eq:cburn}
%\ee

 The other possible limiter of neutronization is the onset of the
explosion. The reactions in Table 1 show that
$Q\approx 16$ MeV is released as thermal energy when six carbon nuclei
are burned.\footnote{Not all of this energy will always go into
heating up the core. For example, if the convective Urca process is
operating, then it will take more energy (and more carbon burning)
to get to sufficient temperatures.} If we let $E_c$ be the total thermal
content that is within the convective core with respect to the initial isothermal
WD,  this implies a change $\Delta Y_e=-\eta E_cm_p/QM_c$
in a convective core of mass $M_{c}$,
\be
	\Delta Y_e = -6.5 \times10^{-4}\frac{\eta}{2}\frac{E_c}{10^{49}\ {\rm ergs}}
	\frac{M_\odot}{M_c},
\ee
For this to compete with the $^{22}$Ne contribution, a total
energy
\be
	E_c = 1.4\times10^{49}\ {\rm ergs}\ \frac{2}{\eta}\frac{X(^{22}{\rm Ne})}{0.02}
			\frac{M_{\rm c}}{M_\odot},
	\label{eq:ec}
\ee
or $7\times10^{15}\ {\rm ergs \ g^{-1}}$,
must be released prior to the explosion.

Simmering ends when dynamical burning is triggered, requiring
$T_c\approx8\times 10^8\ {\rm K}$ \citep{woo04}. If the burning
occurred within a single zone with the specific heat of \S 2, then
reaching this $T_c$ would require $\approx1.3\times10^{16}\ {\rm ergs \ g^{-1}}$,
in excess of that implied by equation (\ref{eq:ec}). Of course, in reality
the convective zone extends outward, so that little
mass is at $T_c$. To accurately determine the resulting neutronization,
we construct hydrostatic WD models consisting of fully convective
cores as described at the beginning of \S \ref{sec:rates}.
We consider isothermal temperatures of either $10^8\ {\rm K}$
or $2\times 10^8\ {\rm K}$. At any given moment
there is a well defined $M_c$ \citep{les06,pir07}, and we evaluate the current thermal
content by integrating the specific heat relative to the initially isothermal WD,
\be
	E_c=\int_0^{M_c}c_p[T(M)-T_i]dM,
\ee
where $T_i$ is the isothermal WD temperature.
In this way we use our time independent models to
find the fraction of carbon that must have burned, $f$, and the
associated $\Delta Y_e$ as $T_c$ and $M_c$ increase with time. We
assume no neutrino losses and thus all $\approx16\ {\rm  MeV}$
of thermal energy contributes to heating.

\begin{figure}
\epsscale{1.2}
\plotone{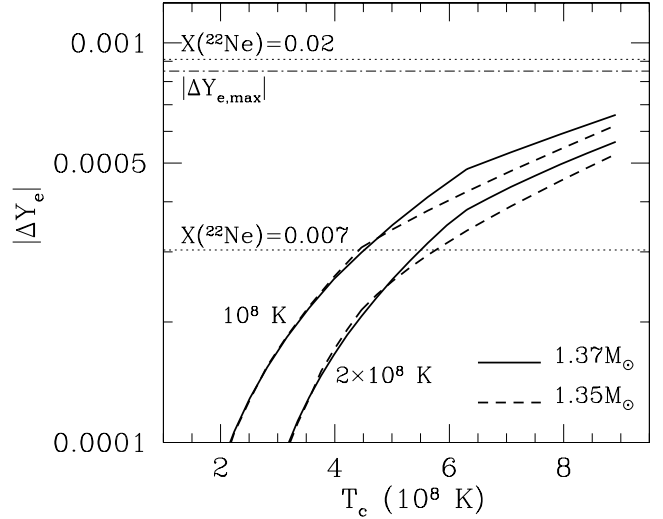}
\caption{Thick dashed ($M=1.35M_\odot$) and solid
($M=1.37M_\odot$) lines show the change of $Y_e$ as a function of the current central
temperature, $T_c$, due to carbon burning during simmering. In each case we consider
isothermal temperatures of $10^8\ {\rm K}$ and $2\times10^8\ {\rm K}$ (as labeled).
The dot-dashed line is the maximum change possible if proton captures on carbon
are only limited by competition from freshly made heavy elements (eq. [\ref{eq:yemax}]).
Dotted lines bracket the change in $Y_e$ expected for $X(^{22}$Ne$)=0.007-0.02$. }
\label{fig:ye}
\epsscale{1.0}
\end{figure}

  In Figure \ref{fig:ye} we summarize the results of these calculations.
In each case, the slope of $\Delta Y_e$ shows a break at the transition
from $\eta=2$ ($t_h>t_{\rm ec,23}$) to $\eta=1$. This break occurs later
for more massive WDs (Fig. \ref{fig:simmering}), thus these have more
neutronization during simmering. Increasing the isothermal temperature
decreases $M_c$, so that it takes less burning to reach a given $T_c$.
These fully integrated models make it clear that substantial neutronization
occurs prior to the explosion. In comparison to the $\Delta Y_e$
from $^{22}$Ne, simmering effects dominate if $X(^{22}$Ne$)<0.013$ or $Z/Z_\odot\lesssim2/3$.
This thwarts the occurrence of high $Y_e$ SNe Ia in low metallicity
progenitors.

% -----------------------------------------------------------
% -----------------------------------------------------------

\section{Conclusion and Discussion}
\label{sec:conclusion}

  We have found that significant neutronization of the WD core occurs throughout the
simmering stage of carbon burning until the onset of the explosion. If substantial
energy is lost to the convective Urca process \citep[][and references therein]{les05},
then the neutronization is truncated by proton captures onto freshly synthesized heavy
elements (resulting in eq. [\ref{eq:yemax}]). The main consequence is a
uniform ``floor'' to the amount of neutronization that dominates over the metallicity dependent
contribution for all progenitors with $Z/Z_\odot\lesssim2/3$.

   Given the likely
significance this has for SNe Ia, more work needs to be done.
In particular, at high ignition densities, heavy element
electron captures and a full reaction network are needed to follow
the resulting diverse collection of elements (see the discussion in \S 2.2).
The convective Urca process %(which also occurs at these high densities)
is another complication we have not addressed. In principle, if more energy is lost
to neutrinos then more burning (and thus more neutronization) is required to make
the burning dynamical. Assessing this will
necessitate coupling a full nuclear network \citep{cha07b}
to convective calculations. Such models would accurately determine
$\eta$ rather than simply setting it to 1 or 2.

In closing, we highlight some important features exhibited by recent
observations of SNe Ia. It is clear that the
amount of $^{56}$Ni produced in SNe Ia has a
dynamic range ($0.1-1M_\odot$) larger than can be
explained by metallicity or simmering neutronization.  However, since an
intriguing trend is the prevalence of $^{56}$Ni rich events in star-forming
regions it is interesting to quantitatively explore how large the observed
discrepancy is.
 Using the SNLS sample of Sullivan et al. (2006), Howell et
al. (2007) found that the average stretch is $s=0.95$ in passive
galaxies (e.g. E/S0's) and $s=1.05$ in star-forming galaxies.
%Howell
%(2007; priv. communication) gives a direct translation to $^{56}$Ni
%masses of $0.51M_\odot$ for $s=0.95$ and $0.64M_\odot$ for
%$s=1.05$.
Using Jha et al's (2006) $\Delta M_{15}(B)-s$ relation and
Mazzali et al.'s (2007) relation between $\Delta M_{15}(B)$ and
$^{56}$Ni mass we get $0.58M_\odot$ ($s=0.95$) and $0.72M_\odot$
($s=1.05$). Hence, amongst the large diversity, there is a tendency
for SNe in star-forming galaxies to produce $\approx 0.13M_\odot$ more
$^{56}$Ni than those in large ellipticals.

 Since the SN Ia rate scales with mass in ellipticals and
star formation rate in spirals (Mannucci et al. 2005; Scannapieco
\& Bildsten 2005; Sullivan et al. 2006), SNe from passive
galaxies in the SNLS survey are from more massive galaxies than
the SNe in star-forming galaxies (Sullivan et al. 2006). Using the
mass-metallicity relation of Tremonti et al. (2004), our integration
of the separate samples in Sullivan et al. (2006) yield average
$12+\log({\rm O/H})=8.87$ in active galaxies and $9.1$ in ellipticals
(solar value is $12+\log({\rm O/H})=8.69$). Due to the increase of
${\rm N/O}$ at high metallicities (Liang et al. 2006), the SNe in ellipticals
have twice as much $^{22}$Ne content as those in spirals. From the
relation of Timmes et al. (2003), this implies $\approx
5\%$ less $^{56}$Ni, whereas the observed decrement is
$>15\%$. Explaining the observed decrement
would require a contrast of $\Delta X(^{22}$Ne$)\approx 0.06$, or
nearly 3 times solar. Although we have found that simmering enhances neutronization,
the effect is not great enough ($\Delta Y_{e,\rm max}$ would give the same change in
neutronization as doubling a solar metallicity),
and a diverse set of core conditions would still be required. A large enhancement could
be present in the core if substantial gravitational separation had
occurred \citep{bh01,db02}, yet convective mixing during simmering
will reduce it based on the fraction of the star that is convective.
For a convection zone that extends out to $M_\odot$, the resulting $^{22}$Ne
enhancement would be at most $\approx 30\%$.

\acknowledgements
We thank E. Brown, R. Ellis, F. Forster, A. Howell, B. Paxton, P. Podsiadlowski, H. Schatz
and F. Timmes for discussions, and D. Yakovlev for sharing carbon ignition curves.
We also thank the referee for constructive comments and questions.
This work was supported by the National Science Foundation
under grants PHY 05--51164 and AST 02-05956.

% -----------------------------------------------------------
% -----------------------------------------------------------

\end{document}